\documentclass{jfm}
\usepackage{graphicx}
\usepackage{epstopdf, epsfig}

\usepackage{amsmath}
\usepackage{float}
\usepackage{bm}

\shorttitle{Guidelines for authors}
\shortauthor{V. \v{S}kult\'ety, A. Morozov}

\title{A note on forces exerted by a Stokeslet on confining boundaries}

\author{Viktor \v{S}kult\'ety
	\and
	Alexander Morozov\corresp{\email{alexander.morozov@ed.ac.uk}}	
}

\affiliation{
	SUPA, School of Physics and Astronomy, The University of Edinburgh, James Clerk Maxwell Building, Peter Guthrie Tait Road, Edinburgh EH9 3FD, United Kingdom	
}

\begin{document}
	
	\maketitle
	
\begin{abstract}
We consider a stokeslet applied to a viscous fluid next to an infinite, flat wall, or in-between two parallel walls. We calculate the forces exerted by the resulting flow on the confining boundaries, and use the results obtained to estimate the hydrodynamic contribution to the pressure exerted on boundaries by force-free self-propelled particles.
\end{abstract}


\section{Introduction}

Solutions to the Stokes equation can be constructed by combining suitably placed Stokelets (the Green function of the Stokes equation) and other singular solutions, that simultaneously satisfy the equation of motion and the boundary conditions \citep{Happel83}. This approach has proven especially fruitful in describing the motion of small solid bodies  \citep{Chwang1975} and self-propelled particles \citep{LaugaPowers2009,Spagnolie2012}. Recent debate on the pressure exerted by microswimmers on the walls of the enclosing container 
\citep{Yang2014,Takatori2014,Solon2015}, together with the observations of the apparent viscosity of microswimmer suspensions being strongly affected by their presence \citep{Lopez2015}, stresses the need to evaluate the forces exerted by microswimmers on solid boundaries. Since their hydrodynamic fields can be constructed from the fundamental solutions of the Stokes equation, it is sufficient to consider the forces due to the latter.

\begin{figure}
	\centering
	\includegraphics[width=12cm]{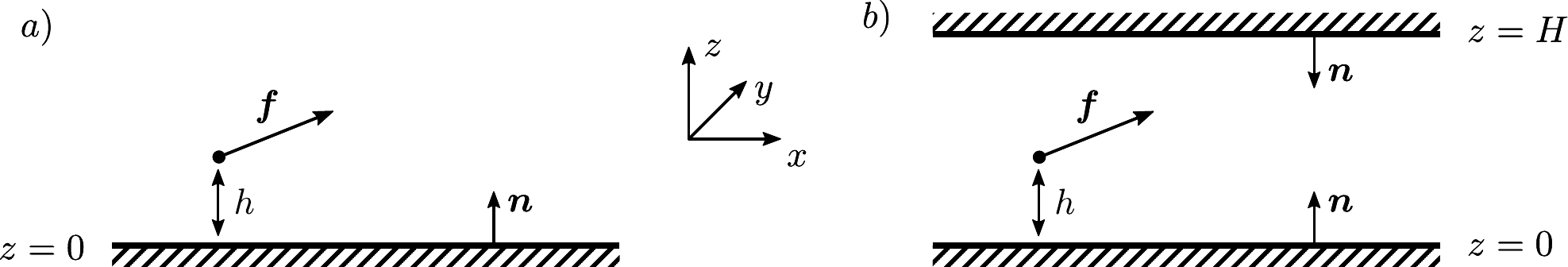} \label{S2:fig.1}
	\caption{Geometries used in this note. a) A point force applied to a fluid next to a wall. b) A point force applied in-between two parallel walls. The unit vectore $\bm n$ gives the direction of the outer normal to each boundary.}
\end{figure}

Here, we study two archetypal problems: a Stokeslet next to a single flat boundary, and a Stokeslet confined in-between to parallel walls, see Fig.\ref{S2:fig.1}. Both problems are solved in a Cartesian coordinate system $\{x,y,z\}$, with the $z$-direction selected perpendicular to the boundaries. The velocity field ${\bm v}({\bm r})$ at a position $\bm r$ satisfies the incompressible Stokes equation
\begin{align}
&-\partial_i p({\bm r}) + \mu \partial^2 v_i({\bm r}) + f_i \delta\left({\bm r} - {\bm r_0}\right) = 0, \label{Stokes1}\\
&\qquad\qquad\qquad\partial_i v_i({\bm r}) = 0, \label{Stokes2}
\end{align}
where $p$ is the pressure, and $\mu$ is the viscosity of the fluid; $\partial_i$ denotes the spatial derivative in the $i$-th direction, $i=\{x,y,z\}$, while $\partial^2$ denotes the Laplacian. The point force $\bm f$ is applied to the fluid at a position ${\bm r_0}$, which, without loss of generality, is chosen to be $\left(0,0,h\right)$. The fluid is assumed to satisfy the no-slip condition  at all boundaries. The solution to Eqs.\eqref{Stokes1} and \eqref{Stokes2} has been obtained by \cite{Blake1971}, for the case of a single boundary, and by \cite{Liron1976} and \cite{Daddi2018}, for two confining walls. Here, we use these results to evaluate the associated forces applied by the fluid on the enclosing boundaries.

\section{A point force next to a single boundary}
\label{Section:Blake}

In this problem, we consider a semi-infinite fluid bounded by an infinite, flat solid boundary at $z=0$ (see Fig.\ref{S2:fig.1}a). The solution to Eqs.\eqref{Stokes1} and \eqref{Stokes2} in this case has been obtained by \cite{Blake1971}, and reads
\begin{align}
&\qquad\qquad v_j = \frac{f_k}{8\pi\mu}\left[ \left(\frac{1}{r}-\frac{1}{R} \right)\delta_{jk} + \frac{r_j r_k}{r^3} - \frac{R_j R_k}{R^3} \right. \nonumber \\
&\left. +2h\left(\delta_{k1}\frac{\partial}{\partial R_1} + \delta_{k2}\frac{\partial}{\partial R_2} -\delta_{k3}\frac{\partial}{\partial R_3}\right )
\left\{ \frac{h R_j}{R^3} - \frac{\delta_{j3}}{R} - \frac{R_j R_3}{R^3} \right\} \right], 
\label{vBlake}\\
&p= \frac{f_k}{8\pi\mu}\left[ \frac{r_k}{r^3} - \frac{R_k}{R^3} 
-2h\left(\delta_{k1}\frac{\partial}{\partial R_1} + \delta_{k2}\frac{\partial}{\partial R_2} -\delta_{k3}\frac{\partial}{\partial R_3}\right ) \left(\frac{R_3}{R^3}\right)
\right],
\label{pBlake}
\end{align}
where ${\bm r}=\left(x,y,z-h\right)$ and ${\bm R}=\left(x,y,z+h\right)$. An infinitesimal force exerted on the boundary by this velocity field is given by \cite{LandauLifshitz}
\begin{align}
dF_i = \Sigma_{ij} \Big\vert_{z=0} n_{j} dx dy, 
\end{align}
where $\bm n$ is the outer normal to the solid boundary, and $\Sigma_{ij}$ is the stress tensor
\begin{align}
\Sigma_{ij} = - p \delta_{ij} + \mu\left(\partial_{j}v_{i} + \partial_{i}v_{j}\right).
\end{align}
Using $n_j = \delta_{jz}$, we obtain for the total force on the boundary
\begin{align}
F_{x} &= \mu \int_{-\infty}^{\infty} dx dy \ \partial_{z} v_{x} \Big\vert_{z=0} \label{ForceX}, \\
F_{z} &= - \int_{-\infty}^{\infty} dx dy \ p \Big\vert_{z=0},
\label{ForceZ}
\end{align}
where we used $\partial_{x} v_{z} \Big\vert_{z=0} = \partial_{y} v_{z} \Big\vert_{z=0} = 0$, since the operations of taking a derivative with respect to $x$ or $y$ and evaluating these velocity components at $z=0$ commute, and $v_i$ vanish at the boundary. In a similar fashion, we set $\partial_{z} v_{z} \Big\vert_{z=0}=0$ in  Eq.\eqref{ForceZ}, which follows from the incompressibility condition, Eq.\eqref{Stokes2}, and the argument above. The expression for the $y$-component of the force is obtained by replacing the subscripts $x$ with $y$ in Eq.\eqref{ForceX}.

Explicit evaluation using Eqs.\eqref{vBlake} and \eqref{pBlake} yields
\begin{align}
&\mu \partial_{z} v_{x} \Big\vert_{z=0}  = \frac{3 h x}{2\pi}\frac{f_x x + f_y y - f_z h}{\left(h^2+x^2+y^2 \right)^{5/2}}, \\
&p \Big\vert_{z=0} =  \frac{3h^2}{2\pi}\frac{f_x x + f_y y - f_z h}{\left(h^2+x^2+y^2 \right)^{5/2}},
\end{align}
which, upon integration in Eqs.\eqref{ForceX}  and \eqref{ForceZ}, give
\begin{align}
F_i = f_i.
\label{ForceBlake}
\end{align}

\section{A point force in a plane channel}

In the second problem we consider a fluid confined in-between two infinite parallel walls placed at $z=0$ and $z=H$ (see Fig.\ref{S2:fig.1}b).  The flow field $v_i({\bm r})$ satisfies Eqs.\eqref{Stokes1} and \eqref{Stokes2} with the boundary conditions $v_i(z=0)=v_i(z=H)=0$. The solution to this problem was first reported by \cite{Liron1976}, who used a method similar to that of \cite{Blake1971}. An alternative approach was developed by \cite{Bickel2007}, and by Daddi-Moussa-Ider and co-workers \citep{Daddi2018b,Daddi2018}, which is more convenient for evaluating the force applied to the boundaries. In what follows, we use the method of \cite{Daddi2018}, and repeat the main steps of their derivation for completeness. Since the result of \cite{Liron1976} is probably better known, in Appendix \ref{appendixA} we repeat the same derivation using their method.

We start by introducing a two-dimensional Fourier transform for the velocity
\begin{align}
v_{i}(x,y,z) = \frac{1}{\left(2\pi\right)^2} \int_{-\infty}^{\infty} d k_x d k_y \ e^{i\left( k_x x + k_y y\right)} \hat{v}_{i}(k_x,k_y,z),
\label{FT}
\end{align}
and a similar transform for the pressure. Upon inserting these expressions into Eqs.\eqref{Stokes1} and \eqref{Stokes2}, we obtain
\begin{align}
-ik_{\alpha} \hat{p} + \mu(\partial_{z}^{2} - k^{2}) \hat{v}_{\alpha} &+ f_{\alpha} \delta(z-h) = 0,
\label{S3:eq.Stokes1} \\
-\partial_{z} \hat{p} + \mu(\partial_{z}^{2} - k^{2}) \hat{v}_{z} &+ f_{z} \delta(z-h) = 0,
\label{S3:eq.Stokes2} \\
i k_{x} \hat{v}_{x} + i k_{y} \hat{v}_{y} &+ \partial_{z} \hat{v}_{z} = 0,  \label{S3:eq.Stokes3}
\end{align}
where $\alpha = \{x,y\} $,  and $ k^2 = k_x^2 + k_y^2$. To proceed, we introduce the longitudinal and transverse components of the in-plane velocity
\begin{align}
\hat{v}_{x} &= \frac{k_{x}}{k} \hat{v}_{l} + \frac{k_{y}}{k} \hat{v}_{t}, \quad \hat{v}_{y} = \frac{k_{y}}{k} \hat{v}_{l} - \frac{k_{x}}{k} \hat{v}_{t}, 
\label{transform}
\end{align}
and a similar transformation for the longitudinal $ f_{l} $ and transverse $f_{t} $ components of the point force. Applying this transformation to Eqs.\eqref{S3:eq.Stokes1}-\eqref{S3:eq.Stokes3}, we obtain
\begin{align}
& \qquad\quad \mu (\partial_{z}^{2} - k^{2}) \hat{v}_{t} + f_{t} \delta(z-h) = 0, \label{S3:eq.vt} \\
& \,-ik\hat{p} + \mu (\partial_{z}^{2} - k^{2}) \hat{v}_{l} + f_{l} \delta(z-h) = 0, \label{S3:eq.vl} \\
& -\partial_{z} \hat{p} + \mu(\partial_{z}^{2} - k^{2}) \hat{v}_{z} + f_{z} \delta(z-h) = 0, \label{S3:eq.vz} \\
& \qquad\qquad\qquad i k \hat{v}_{l} + \partial_{z}\hat{v}_{z} = 0. \label{S3:eq.CE}
\end{align}
These equations de-couple the transverse component from the rest, and below we solve the associated problems separately.

\subsection{Transverse velocity component}

To solve Eq.\eqref{S3:eq.vt}, we observe that its solution can be split into two parts, $\hat{v}_{t}^{+}$ for $z>h$, and $\hat{v}_{t}^{-}$ for $z<h$, that satisfy that same equation
\begin{align}
\mu (\partial_{z}^{2} - k^{2}) \hat{v}_{t}^{\pm} &= 0, \label{S3:eq.vt.1}
\end{align}
and the boundary conditions
\begin{align}
\hat{v}_{t}^{-}(0) = \hat{v}_{t}^{+}(H) = 0.
\end{align}
The matching condition at $z=h$ is obtained by integrating Eq.\eqref{S3:eq.vt} from $z=h-\varepsilon$ to $z=h+\varepsilon$, which, in the limit $\varepsilon\rightarrow 0$, yields 
\begin{align}
\partial_{z} \hat{v}_t^{+}(h) - \partial_{z} \hat{v}_t^{-}(h) = - f_{t}/\mu. \label{S3:eq.vt.3}
\end{align}
Together with the requirement that the velocity is continuous, $v_{t}^{+}(h)=v_{t}^{-}(h)$, this fully specifies the solution, which is given by
\begin{align}
\hat{v}_{t}^{-}(z) &= \frac{f_{t}}{k \mu} \frac{\sinh(k(H-h))}{\sinh(kH)} \sinh(kz), \label{S3:eq.vt.4} \\
\hat{v}_{t}^{+}(z) &= \frac{f_{t}}{k \mu} \frac{\sinh(kh)}{\sinh(kH)} \sinh(k(H-z)). \label{S3:eq.vt.5}
\end{align}

\subsection{Longitudinal and vertical  velocity components}

Excluding the pressure from Eqs.\eqref{S3:eq.vl} and \eqref{S3:eq.vz}, and using the incompressibility condition, Eq.\eqref{S3:eq.CE}, we obtain for the vertical velocity
\begin{align}
\mu (\partial_{z}^{2} - k^{2})^2 \hat{v}_{z} -  k^{2} f_{z}  \delta(z-h) - ik f_{l} \partial_{z} \delta(z-h) &= 0. \label{S3:eq.vz.1}
\end{align}
Similar to the transverse case, this equation is solved by splitting its solution into two components, $\hat{v}_{z}^{\pm}$, that satisfy
\begin{align}
& \qquad\qquad (\partial_{z}^{2} - k^{2})^2 \hat{v}_{z}^{\pm} = 0, \label{S3:eq.vz2}\\
& \hat{v}_{z}^{-}(0) = \partial_{z} \hat{v}_{z}^{-}(0) = \hat{v}_{z}^{+}(H) = \partial_{z} \hat{v}_{z}^{+}(H) = 0.  \label{S3:eq.vz.BC}
\end{align}
Repeated integration of Eq.\eqref{S3:eq.vz.1} in a small vicinity of $z=h$ yields the following matching conditions
\begin{align}
\partial_{z}^{3} \hat{v}_{z}^{+}(h) - \partial_{z}^{3} \hat{v}_{z}^{-}(h) &= k^{2} \mu^{-1} f_{z}, \label{S3:eq.vz.BC1} \\
\partial_{z}^{2} \hat{v}_{z}^{+}(h) - \partial_{z}^{2} \hat{v}_{z}^{-}(h) &= ik\mu^{-1} f_{l}, \\
\partial_{z} \hat{v}_{z}^{+}(h) - \partial_{z} \hat{v}_{z}^{-}(h) &= 0, \\
\hat{v}_{z}^{+}(h) - \hat{v}_{z}^{-}(h) &= 0. \label{S3:eq.vz.BC2}
\end{align}
The solution to Eqs.\eqref{S3:eq.vz2}-\eqref{S3:eq.vz.BC2} is given by
\begin{align}
v_{z}^{\pm} &= \frac{T_{zz}^{\pm} f_{z} + T_{zl}^{\pm} i k f_{l}}{4 k \mu  \left(1 + 2 H^2 k^2 - \cosh (2 H k)\right)}, \label{S3:eq.vz.4} 
\end{align}
where 
\begin{align}
T_{zz}^{-}(z,h) =& -\left(2 H k^2 (-h+H+z)+1\right) \sinh (k (h-z)) \nonumber \\
&\ +k \left(z \left(4 H k^2 (h-H)-1\right)+h-2 H\right) \cosh (k (h-z)) \nonumber \\
&\ +\left(2 k^2 (H-h) (H-z)+1\right) \sinh (k (h+z)) \nonumber \\
&\ - k(h-2 H+z) \cosh (k (h+z)) \nonumber \\
&\ -\left(2 h k^2 z+1\right) \sinh (k (h-2 H+z)) \nonumber \\
&\ + k(h+z) \cosh (k (h-2 H+z)) \nonumber \\
&\ +\sinh (k (h-2 H-z)) \nonumber \\
&\ +k (z-h) \cosh (k (h-2 H-z)), \\
T_{zl}^{-}(z,h) =& \left(4 H k^2 z (H-h)+(z-h)\right) \sinh (k (h-z)) \nonumber \\
&\ -2 H k (h-H+z) \cosh (k (h-z)) \nonumber \\
&\ +(h-z) \sinh (k (h+z)) \nonumber \\
&\ +2 k (h-H) (H-z) \cosh (k (h+z)) \nonumber \\
&\ +(z-h) \sinh (k (h-2 H+z)) \nonumber \\
&\ +2 h k z \cosh (k (h-2 H+z)) \nonumber \\
&\ +(h-z) \sinh (k (h-2 H-z)), 	
\end{align}
and 
\begin{align}
T_{zz}^{+}(z,h) &= T_{zz}^{-}(H-z,H-h), \\
T_{zl}^{+}(z,h) &= -T_{zl}^{-}(H-z,H-h).
\end{align}
The longitudinal component $\hat{v}_{l}$ can now be obtained from Eq.\eqref{S3:eq.CE}, while the pressure is given by Eq.\eqref{S3:eq.vl}.

\subsection{Forces exerted on the boundaries}
\label{sub:Forces}

The forces applied by the flow determined above can now be calculated in a manner similar to Section \ref{Section:Blake},
and are given by
\begin{align}
F_{x}^{\pm} &= \mp \mu \int_{-\infty}^{\infty} dx dy \ \partial_{z} v_{x}^{\pm}, \label{ForceX0H} \\
F_{z}^{\pm} &= \pm \int_{-\infty}^{\infty} dx dy \ p^{\pm}, \label{ForceZ0H}
\end{align}
evaluated at $z=H$ and $z=0$, respectively. Here, $v_x^{\pm}$ and $p^{\pm}$ are the inverse transforms of the corresponding Fourier components, and we used the fact that the outer normal at the $z=H$ boundary is pointing in the negative $z$-direction. The integrals in Eqs.\eqref{ForceX0H} and \eqref{ForceZ0H} can, in fact, be obtained from the Fourier transform introduced in Eq.\eqref{FT}. Indeed, if we put $k_x = k \cos{\theta}$ and $k_y = k \sin{\theta}$, 
for an arbitrary function $\phi$ that depends on $x$ and $y$ in a symmetric manner we obtain
\begin{align}
\int_{-\infty}^{\infty} dx dy \ \phi(x,y) = \frac{1}{2\pi} \lim\limits_{k \rightarrow 0} \int_{0}^{2\pi} d\theta \ \hat\phi(k,\theta).
\end{align}
Therefore, the forces on the boundaries are readily obtained by integrating $\mp \mu \partial_{z} \hat{v}_{x}^{\pm}$ and $\pm \hat{p}^{\pm}$ over $\theta$, taking the limit $k\rightarrow 0$, and evaluating the result at the appropriate $z$.
The final results then read
\begin{align}
F_{\alpha}^{-} &= (1-\Delta)(1-\frac{3}{2}\Delta) f_{\alpha}, \qquad F_{z}^{-} = (1-\Delta)^{2}(1+2\Delta) f_{z}\label{ForceChannelMinus}\\
F_{\alpha}^{+} &= \frac{1}{2}\Delta(3\Delta - 1) f_{\alpha},\qquad\qquad F_{z}^{+} = \Delta^{2} (3-2\Delta) f_{z} 
\label{ForceChannelPlus}
\end{align}
where $ \Delta = h/H $, and $\alpha=\{x,y\}$.

\section{Discussion}

Equations \eqref{ForceBlake} and \eqref{ForceChannelMinus}-\eqref{ForceChannelPlus} constitute the main results of this work. The first case corresponds to a Stokeslet near an infinite plane wall and implies that the whole force applied to the fluid is transmitted to the wall, \emph{independent} of the Stokeslet's distance to the wall. While appearing surprising, this result can be understood from a simple argument. Due to the linearity of the Stokes equation, we expect the force on the wall to be proportional to the strength of the force applied to the fluid, $F_i = g(h) f_i$, where $g(h)$ is an unknown function of the distance between the Stokeslet and the wall. Since $g(h)$ should be dimensionless, it can only depend on a ratio between $h$ and another length-scale. However, there are no other length-scales in the problem, and $g$ is constant, independent of $h$. Considering the case when the Stokeslet is applied directly to the interface between the wall and the fluid fixes $g=1$, giving the result in Eq.\eqref{ForceBlake}. An interesting consequence of this result is that an arbitrary force distribution applied to the fluid next to a single wall exerts no force on the wall, as long as the total force applied to the fluid is zero, as in the case of a collection of force-free self-propelled particles. In a similar fashion, a force-free microswimmer stalled by the wall, exerts no total force on it. Indeed, the propulsive force generated by the swimmer is directly transmitted to the wall through the action of the interaction potential between the wall and the swimmer. To generate this propulsive force, the swimmer applies the equal and opposite force on the fluid some distance away from the wall, which is fully transmitted to the wall, as Eq.\eqref{ForceBlake} suggests. The total sum is zero for any orientation of the swimmer in contact with the wall. Therefore, there is no hydrodynamic contribution to the pressure from  a suspension of force-free swimmers next to a single boundary.

When the Stokeslet is confined between two parallel walls, the argument above yields $F_i = g(h/H) f_i$, since there are now two length-scales in the problem. The corresponding functions $g$ are non-trivial and different for the force components perpendicular and parallel to the wall, see Eqs.\eqref{ForceChannelMinus} and \eqref{ForceChannelPlus}. 
First, we observe that these expressions are symmetric with respect to $\Delta\rightarrow 1-\Delta$, as expected.
Next, in the limit of $H\rightarrow\infty$, keeping $h$ finite, we recover Eq.\eqref{ForceBlake} for the force on the lower wall, while $F_{i}^{+}=0$; the same holds for $h\rightarrow\infty$, keeping $H-h$ finite, with $F_{i}^{+}=f_i$ and $F_{i}^{-}=0$. Finally, the correct behaviour is also recovered in the limits of $h\rightarrow0$ and $h\rightarrow H$. 

Eqs.\eqref{ForceChannelMinus} and \eqref{ForceChannelPlus} also allow us to make an interesting observation regarding the total force applied to both boundaries. While the total vertical force on the walls is equal to the vertical force applied to the fluid, $F_{z}^{-} +F_{z}^{+} = f_z$, the horizontal components give $F_{\alpha}^{-} +F_{\alpha}^{+} = (1-3\Delta(1-\Delta)) f_{\alpha}$, with $\alpha=\{x,y\}$. The latter result implies that $F_{\alpha}^{-} +F_{\alpha}^{+} \le f_i$, where the equality only applies when $\Delta=0$ or $1$. To understand the origin of the 'missing' force, we consider an imaginary box around the Stokeslet and calculate the forces applied to the planes $x=\pm L$ and $y=\pm L$, where $L\gg H$. Far away from the Stokeslet, the velocity field is given by Eq. (51) of \cite{Liron1976}, and has only the in-plane components, while the far-field behaviour of the pressure can be deduced from Eq.(56) of the same reference. Calculating the forces exerted by this velocity field in the $x$-direction on the fictitious surfaces as $L\rightarrow\infty$, we obtain that the forces at $y=\pm L$ are zero, while the forces at $x=\pm L$ are the same and equal to $(3/2)\Delta(1-\Delta)f_x$, where only the pressure term contributes to this result. An identical expression is, of course, obtained for the $y$-component of the force, where only the fictitious surfaces perpendicular to the $y$-axis experience non-zero forces. Together with Eqs.\eqref{ForceChannelMinus} and \eqref{ForceChannelPlus}, this gives the total force applied to the boundaries enclosing the Stokeslet being equal to $f_i$, as it should. 

We conclude by observing that our results can be trivially generalised for an arbitrary distribution of point forces applied to the fluid due to linearity of the Stokes equation. In particular, we consider a force dipole, which is relevant for force-free self-propelled microswimmers \citep{LaugaPowers2009}. The dipole consists of two equal and opposite point forces, $-f \bm{e}$ and $f \bm{e}$, applied to the fluid at $(0,0,h)$ and $(0,0,h)+l \bm{e}$, respectively, where $f$ is the magnitude of the force, $\bm{e}$ is a unit vector along the direction of the dipole, and $l$ is its length. From Eq.\eqref{ForceChannelPlus}, the vertical component force on the upper boundary due to the dipole is given by
\begin{align}
F_d(h) = \frac{f l}{H} \left[ -2 \frac{l^2}{H^2} e_z^4 + 6 e_z^2 \Delta(1-\Delta) +3 e_z^3\frac{l}{H}(1-2\Delta)\right],
\end{align}
where $e_z$ denotes the $z$-component of $\bm{e}$. An equal and opposite force is applied to the lower boundary.
Next, we consider a collection of such dipoles at a number density $n$. Although it has been demonstrated that suspensions of dipolar microswimmers exhibit significant correlations even at low densities \citep{Stenhammar2017}, here we assume the suspension to be homogeneous and isotropic, for simplicity. The pressure on the upper wall (a force per unit area) can then be calculated as the following average 
\begin{align}
p_d = \frac{n}{2} \int_{0}^{\pi}d\theta \sin{\theta} \int_{0}^{H} dh F_d(h) \approx \frac{1}{3} f l n,
\end{align}
where we used $e_z=\cos\theta$ in spherical coordinates, and neglected terms of order $l/H$. Apart from a numerical factor, this result can be readily obtained from dimensional analysis. Using the dipolar strength $f l \sim 8 \cdot 10^{-19}$N$\cdot$s 
as measured by \cite{Drescher2011} for \emph{E.coli} bacteria, and setting $n\sim10^{9}$ml$^{-1}$, as in typical experiments with dilute bacterial suspensions \citep{Jepson2013,Lopez2015}, we obtain $p_d\sim10^{-4}$Pa. Such pressures are too small to be measured by conventional rheometry but, perhaps, can be observed in an appropriate microfluidic experiment. We would like to note that the pressure calculated above is due to the velocity fields generated by the swimmers, and does not contain the osmotic contribution \citep{Yang2014,Takatori2014,Solon2015}.

Discussions with Mike Cates, Wilson Poon, Saverio Spagnolie, and Julien Tailleur are gratefully acknowledged.

\appendix

\section{Alternative derivation of Eqs.\eqref{ForceChannelMinus}-\eqref{ForceChannelPlus} }
\label{appendixA}

Here we demonstrate that the forces exerted on the walls of a plane channel by a Stokeslet can also be derived with the help of the velocity field obtained by \cite{Liron1976}, which is probably the most famous treatment of that problem.

Their solution for the $j$-th component of the velocity field due to the $k$-th component of the point force, $u_{j}^{k} $, is decomposed into two parts
\begin{align}
u^{k}_{j} = v^{k}_{j} + w^{k}_{j},
\end{align}
where $v^{k}_{j}$ is the contribution due to the original free-space Stokeslet, together with an infinite number of its images, and $w^{k}_{j}$ is an auxiliary solution that ensures the no-slip boundary conditions at the walls. 
The Fourier transform of the auxiliary solution is given by Eqs.(26) and (31) of \cite{Liron1976}; note that their Fourier transform convention differs from ours, Eq.\eqref{FT}, by $2\pi$. Using the same argument as in Section \ref{sub:Forces}, we express the contribution of the auxiliary solution to the forces on the upper boundary as
\begin{align}
&F^{+}_{w,\alpha} = - \mu f^k \lim_{\zeta\rightarrow\infty}\int_{0}^{2\pi} d\theta \,\partial_z \hat{w}^{k}_{\alpha}(\lambda_x=\zeta\cos\theta,\lambda_y=\zeta\sin\theta,z=H), \\
&F^{+}_{w,z} = f^k \lim_{\zeta\rightarrow\infty}\int_{0}^{2\pi} d\theta \,\hat{p}^{k}(\lambda_x=\zeta\cos\theta,\lambda_y=\zeta\sin\theta,z=H),
\end{align}
where $\lambda_x$ and $\lambda_y$ are the analogues of $k_x$ and $k_y$ used in the main text, and $\zeta^2=\lambda_x^2+\lambda_y^2$, as in \cite{Liron1976}. Here, $\alpha=\left\{x,y\right\}$, and $\hat{w}^{k}_{\alpha}$ and $\hat{p}^{k}$ denote the Fourier transforms of the auxiliary velocity and pressure, respectively. Performing the integrals and taking the limit yields
\begin{align}
&F^{+}_{w,\alpha} = -\frac{3}{2}\Delta(1-\Delta) f_\alpha, \label{Fwalpha}\\
&F^{+}_{w,z} = -\Delta \left(1-\Delta \right)\left(1 - 2\Delta \right) f_z. \label{Fwz}
\end{align}

The velocity and pressure fields due to the original free-space Stokeslet and its images are given in Eqs.(15) and (16) of \cite{Liron1976}, and are conveniently expressed in terms of the infinite series from Eqs.(43) and (44) \emph{ibid.} Using Eqs.\eqref{ForceX0H} and \eqref{ForceZ0H}, we obtain
\begin{align}
&F^{+}_{v,\alpha} = -\frac{f_\alpha}{2\pi H^2} \int_{-\infty}^{\infty} dx dy \nonumber \\
& \qquad\qquad \times
\sum_{n=1}^{\infty} \left(-1\right)^{n} \pi n \sin{\left(\pi n \Delta\right)} 
\left[ K_0\left( \frac{\pi n \rho}{H}\right) +\frac{x^2}{\rho}\frac{\pi n}{H} K_1\left( \frac{\pi n \rho}{H}\right)
\right]  = f_\alpha \Delta, \\
&F^{+}_{v,z} = -\frac{f_z}{\pi H^2} \int_{-\infty}^{\infty} dx dy \sum_{n=1}^{\infty} \left(-1\right)^{n} \pi n \sin{\left(\pi n \Delta\right)} 
K_0\left( \frac{\pi n \rho}{H}\right) = f_z \Delta,
\end{align}
where $\rho^2 = x^2 + y^2$, $K_0$ and $K_1$ denote the zeroth- and first-order modified Bessel functions of the second kind, and we dropped the terms that do not contribute to the force. Combining these expressions with Eqs.\eqref{Fwalpha} and \eqref{Fwz}, we arrive at Eq.\eqref{ForceChannelPlus}. The force on the lower boundary is obtained by replacing $\Delta$ with $1-\Delta$ in Eq.\eqref{ForceChannelPlus}, as can be seen from Eq.\eqref{ForceChannelMinus}.

\bibliographystyle{jfm}
\bibliography{refs}

\end{document}